\documentclass[fleqn,10pt]{wlscirep}
\usepackage[utf8]{inputenc}
\usepackage[T1]{fontenc}
\usepackage{type1cm}
\usepackage{ae,aecompl}
\usepackage{microtype}
\usepackage{hyperref}
\hypersetup{colorlinks,allcolors=black}

\usepackage{booktabs}
\usepackage{multirow}
\usepackage{multicol}

\usepackage{soul}
\newcommand{\ctext}[3][RGB]{%
  \begingroup
  \definecolor{hlcolor}{#1}{#2}\sethlcolor{hlcolor}%
  \hl{#3}%
  \endgroup
}

\usepackage{graphicx}

\newcommand{\Ni}{(1)~}
\newcommand{\Nii}{(2)~}
\newcommand{\Niii}{(3)~}

\usepackage{xspace}
\newcommand{\webisReuseCorpus}{\texttt{Webis-STEREO-21}\xspace}

\begin{document}

\title{A large dataset of scientific text reuse in Open-Access publications}

\author[1,*]{Lukas Gienapp}
\author[1]{Wolfgang Kircheis}
\author[1]{Bjarne Sievers}
\author[2]{Benno Stein}
\author[1,*]{Martin~Potthast}
\affil[1]{Leipzig University, Text Mining and Retrieval Group, Leipzig, DE-04109, Germany}
\affil[2]{Bauhaus-Universität Weimar, Web Technology and Information Systems Group, Weimar, DE-99423, Germany}
\affil[*]{corresponding authors:Lukas Gienapp~(lukas.gienapp@uni-leipzig.de)
}

\begin{abstract}
We present the \webisReuseCorpus dataset, a massive collection of \textbf{S}cientific \textbf{Te}xt \textbf{Re}use in \textbf{O}pen-access publications. It contains 91~million cases of reused text passages found in 4.2~million unique open-access publications. Cases range from overlap of as few as eight words to near-duplicate publications and include a variety of reuse types, ranging from boilerplate text to verbatim copying to quotations and paraphrases. Featuring a high coverage of scientific disciplines and varieties of reuse, as well as comprehensive metadata to contextualize each case, our dataset addresses the most salient shortcomings of previous ones on scientific writing. The \webisReuseCorpus does not indicate if a reuse case is legitimate or not, as its focus is on the general study of text reuse in science, which is legitimate in the vast majority of cases. It allows for tackling a wide range of research questions from different scientific backgrounds, facilitating both qualitative and quantitative analysis of the phenomenon as well as a first-time grounding on the base rate of text reuse in scientific publications.

\end{abstract}

\maketitle

\section*{Background \& Summary}

The reuse of text has a longstanding history in science. In qualitative research, besides verbatim quotations, the techniques of paraphrasing, translation, and summarization are instrumental to both teaching and learning scientific writing as well as to gaining new scientific insights \cite{sun:2015}. In quantitative research, the use of templates as an efficient way of reporting new results on otherwise standardized workflows is common \cite{anson:2020}. As science often progresses incrementally, authors may also reuse their texts across (different types of) subsequent publications on the same subject (also called ``text recycling'') \cite{moskovitz:2015,hall:2018,anson:2020}. Likewise, in interdisciplinary research, reuse across publications at venues of different disciplines has been observed to better promote the dissemination of new insights \cite{bird:2002,wen:2007}. Independent of all these manifestations of text reuse is the academic context that establishes their legitimacy: In certain circumstances, any of them may be considered plagiarism, i.e., intentional reuse of a text without acknowledging the original source, in violation of the honor code and academic integrity \cite{eberle:2013}.

Text reuse has been quantitatively studied in many scientific disciplines \cite{ganascia:2014,citron:2015,sun:2015,horbach:2019,anson:2020};
yet few studies assess the phenomenon at scale beyond what can be manually analyzed \cite{citron:2015,horbach:2019}. Large-scale studies require the use of automatic text reuse detection technology. This being both algorithmically challenging and computationally expensive, lack of expertise or budget may have prevented such studies. Employing proprietary analysis software or services instead, too, is subject to budgetary limitations, in addition to their lack of methodological transparency and reproducibility.

Text reuse detection itself is still subject to ongoing research in natural language processing and information retrieval. Setting up a custom processing pipeline thus demands an evaluation against the state of the art. The challenges in constructing a competitive solution for this task arise from the aforementioned diversity of different forms of text reuse, the large solution space of detection approaches, and the need to apply heuristics that render a given solution sufficiently scalable. Preprocessing a collection of scientific publications, too, presents its own difficulties. This includes the noisy and error-prone conversion of publications' original PDF versions to machine-readable texts and the collection of reliable metadata about the publications. The available quantitative studies on scientific text reuse lack with respect to the presentation of preprocessing steps taken, the design choices of the solution to text reuse detection, and their justification in terms of rigorous evaluation. Altogether, comparable, reproducible, reliable, and accessible research on the phenomenon of scientific text reuse remains an open problem.

To provide for a solid new foundation for the investigation of scientific text reuse within and across disciplines, we compile \webisReuseCorpus. To overcome the aforementioned issues, we stipulate three design principles for the creation of the dataset:
\Ni
high coverage, both in terms of the number of included publications and the variety of scientific disciplines;
\Nii
a scalable approach to reuse detection with a focus on high precision at a competitive recall, capturing a comprehensive set of reused passages as reliable resource for research on scientific text reuse; and 
\Niii
comprehensive metadata to contextualize each case, to address a wide range of potential hypotheses.

\webisReuseCorpus results from applying scalable text reuse detection approaches to a large collection of scientific open-access publications, exhaustively comparing all documents to extract a comprehensive dataset of reused passages between them. It contains more than 91~million cases of reused passages among 4.2~million unique publications. The cases stem from 46~scientific fields of study, grouped into 14~scientific areas in all four major scientific disciplines, and spanning over 150~years of scientific publishing between~1860 and~2018. The data is openly accessible to be useful to a wide range of researchers with different scientific backgrounds, enabling both qualitative and quantitative analysis.

\section*{Methods}
\label{sec:method}

\begin{figure}[ht]
\centering
\includegraphics[width=\textwidth]{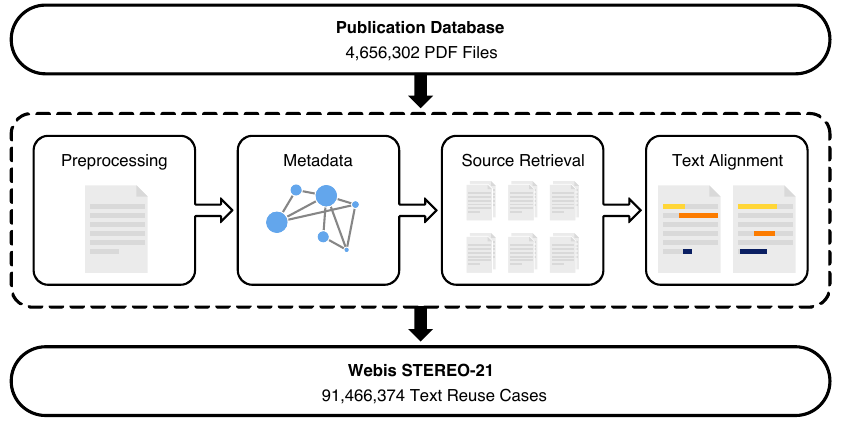}
\caption{Schematic overview of the text reuse detection pipeline. Each document is pre-processed and supplied with metadata. Source retrieval identifies document pairs with local similarities, and alignment is applied to identify reuse cases between those.}
\label{fig:pipeline}
\end{figure}

To create a dataset of text reuse in scientific texts, four things are needed: an operationalization of the phenomenon of text reuse, a large collection of scientific publications, detailed metadata for each of these publications, and a scalable yet effective method for detecting reuse in these publications. In this section we describe all four, explaining the latter both generally and formally to ensure their reproducibility. Figure~\ref{fig:pipeline} illustrates the processing pipeline for the creation of the dataset.

\subsection*{An Operationalization of Text Reuse}

To reuse something means to use it again after the first time. Reused text is text that is primarily, if not exclusively, derived from another text. In academic writing, writing techniques for reusing a text include boilerplate, quotation, paraphrasing, and summarizing. What all these techniques have in common is that a reused text and its original have a certain kind of similarity that the reader can recognize \cite{weber-wulff:2002}. Manual identification of text reuse between two given texts is therefore based on the identification of the relevant text passages where such similarities can be detected.

To operationalize this notion of reuse, a suitable text similarity measure and a similarity threshold have to be chosen, where a pair of texts has to exceed the threshold in order to be considered potential reuse. We broadly distinguish similarity measures that operate at the syntactic level from measures that operate at the semantic level of language, where the former capture the ``reuse of words'' and the latter the ``reuse of ideas'' \cite{sadeghi:2019}.

In practice, text reuse detection relies heavily on syntactic similarity detection. \cite{moskovitz:2021,vroniplag:2021}. Syntactic reuse is relatively easy to visualize and consequently faster to check than semantic reuse \cite{riehmann:2015}. The former can be checked by distant reading, the latter requires close reading, which causes high costs with increasing text length \cite{moretti:2013}. Similarly, it can be assumed that semantic reuse is much less common than syntactic reuse, since the most common goal of text reuse is to save time and cognitive effort, whereas the time savings of semantic reuse are generally lower \cite{potthast:2013c}.

We opt for a conservative similarity analysis at the syntactic level, measuring the correspondence between the surface forms of words occurring in two given texts and the phrases formed from them. This design decision is also motivated by the target domain: In addition to the contribution of new ideas, a large fraction of scientific contributions describe reflection on as well as advances to known ideas, and the development of solutions to tasks and problems up to the point of transfer to practice. A semantic similarity score in this context would rather lead to a citation graph mixing natural matches of ideas with intended reuse. It is in the nature of current semantic similarity measures that they often capture even minor and distant similarities, leading to significant noise in the form of false positives in tasks such as text reuse detection. Also, detection at the syntactic level is more in line with the concept of text reuse used in practice.In other words, `similar' here refers to an overlap in the vocabulary and wording of two documents, an operationalization common to large-scale text reuse detection \cite{martin:2015}.

Although even a single word overlap between two pieces of text, such as a specific and unlikely spelling mistake, can be sufficient as a strong indication of reuse, automatic methods are not yet able to reliably detect such cases. Instead, several overlaps of words and phrases are required that occur in close proximity to each other in both texts, thus forming potentially reused text passages. Such passages are not necessarily verbatim copies: Words and phrases may be added, removed, or changed, and sentences may be rearranged. Nevertheless, sufficient overlap must remain, which can be modeled by the similarity threshold parameter mentioned above. This parameter cannot be derived formally, but must be determined empirically.

This operationalization of text reuse, and thus the cases included in our dataset, are orthogonal to the question of the legitimacy of reuse. Thus, the notion of ``text reuse'' is explicitly not limited to or equated with ``plagiarism.'' We do not draw any conclusions from the fact that two documents have a reused passage in common, but merely observe that there is one. The distinction between original and plagiarized text cannot and should not be an algorithmic decision \cite{weber-wulff:2019}. Further limitations are discussed in the Usage Notes section.

\subsection*{Text Reuse Detection in Large Document Collections}

Detecting all reuses of text in a collection of documents is a problem of high computational complexity. Each document must be compared pairwise with every other document in the collection. So for $N$~documents, $N\cdot (N-1)$~document-to-document comparisons must be made. Each individual comparison requires a considerable amount of time and resources, so that the processing of larger collections of documents exceeds even large computational capacities. In order to still be able to analyze large document collections for text reuse, the set of comparisons is pruned by filtering out document pairs that are guaranteed not to exhibit text reuse according to our operationalization.

This is achieved through a two-step process: first, a cheap-to-compute heuristic is applied to identify candidate pairs of documents where text reuse is likely. Then, the expensive document comparison is performed only for the candidate pairs identified in this way. All other pairs are skipped.The first step of this two-part process is commonly called a source retrieval \cite{stein:2007d}, while the second step is called text alignment \cite{potthast:2013e}. The result of the whole process is a set of cases of text reuse between documents in the collection. A case of text reuse is modeled here as a pair of text passages, one in each document involved in the comparison, that have sufficient overlap of words and phrases, along with references to their source documents and where exactly they are found in them.

While source retrieval greatly improves the efficiency of the overall process, it introduces an error in the form of reduced recall, i.e., pairs of reused passages are overlooked that would have been discovered by an exhaustive comparison of each pair of documents. The goal of source retrieval is therefore to filter out only those document pairs for which the text alignment step will not find any reused passages, which is to be expected for the vast majority of all document pairs. We choose the source retrieval parameters conservatively, so that only those document pairs are skipped for which the text alignment is guaranteed to return no result.

\subsection*{Publication Acquisition, Preprocessing and Metadata}\label{sub:data}

Detecting scientific text reuse and contextualizing it on a large scale requires a large collection of scientific publications and detailed metadata about them. We compiled such a dataset in the five steps of document selection, plain text extraction, text preprocessing, metadata acquisition, and metadata standardization.

We build on the CORE dataset \cite{knoth:2012}, one of the largest collections of open-access scientific publications sourced from more than 12,000~data providers. First, we identify the 6,015,512~unique open-access~DOIs in the March~1,~2018 CORE dataset. Since the plain texts extracted from the PDF files of the publications, as provided by the CORE data, are of varying quality and no structural annotations (such as markup for citations, in-text references, section annotations) are available, we chose to obtain the original PDF~files of the identified open-access~DOIs from various publicly available repositories.

The plain text extraction has been repeated on the acquired PDF~files using the standardized state-of-the-art toolchain GROBID \cite{lopez:2015}. A minimum of~1,000 and a maximum of 60,000~space-separated words are introduced as an effective heuristic to filter out common plaintext extraction errors. In total, we obtained and extracted clean plaintext for 4,267,166~documents (70\%~of the open access publications in the original CORE~dataset). Since text alignment analyzes word overlaps between documents, the highly standardized meta-information in scientific texts such as citations, numbers, author names, affiliations, and references lead to exponentially more false-positive detections of reuse than without them. Therefore, text alignment processes the abstract and main body of a document without in-text references, tables, figures, bibliographic data, and numeric data, while normalizing or removing special characters, which minimizes the false positives observed in preliminary studies. However, research on text reuse and its evaluation on a case-by-case basis benefits from, or even requires, the in-text metadata mentioned above, so that two versions of each text passage involved in a reuse case were kept: the one that came from GROBID, including the aforementioned information except for tables and images, and, the one that was fed into the text alignment.

Based on the detections, we augment the metadata provided by CORE with additional data from the Microsoft Open Academic Graph~(OAG) \cite{tang:2008,sinha:2015}, which contains study field annotations for a large number of publications. Metadata is assigned by matching records in~CORE and~OAG using an article's DOI~identifier. Since the annotated disciplines in the~OAG do not follow a hierarchical scheme and since they are of different granularity per publication (e.g., ``humanities'' as a whole vs. ``chemical solid state research'' as a subfield of chemistry), we manually map the classification found in the~OAG to the standard hierarchical \textit{DFG~Classification of Scientific Disciplines, Research Areas, Review Boards and Subject Areas} \cite{dfg:2016}. We have chosen to replace the term ``review board'' used in the DFG~classification with the more conventional term ``field of study''. The mapping was done by three people independently. In the few cases where there was disagreement, consensus was reached through discussion. The final mapping includes 46~individual scientific fields of study drawn from 14~scientific areas and four major scientific disciplines.

\subsection*{Source Retrieval}

An exhaustive comparison of all four million publications greatly exceeds the available computing capacity. Therefore, a source retrieval step has to be carried out first \cite{hagen:2015e,hagen:2017e}. By applying suitable heuristics, the source retrieval step can be operationalized with linear time complexity with respect to the document count \cite{stein:2019c}, compared to the quadratic expenditure of an exhaustive comparison.

The source retrieval component of our pipeline, i.e., the computation of all candidate comparisons~$D_d$ for a given document~$d$, is operationalized by treating text reuse as a locally bounded phenomenon. Candidate pairs are presumed to be identifiable by comparing text passages between two documents. If the similarity is sufficiently high for at least one combination of passages between two documents~$d$ and~$d'$, then~$d'$ is considered a candidate for~$d$. 

To achieve high scalability, we implement the similarity computation by applying locality-sensitive hash functions~$h$ to each passage~$t$ in a document~$d$, thus representing each document's passages with a set of hash values~$h(t)$. These can be interpreted as a set of fingerprints, encoding the characteristics of each passage in a document in compressed form. The similarity between two passages~$t$ and~$t'$ stemming from two different documents~$d$ and~$d'$ is then approximated by the extent of overlap between their hash sets,~$|h(t)\cap h(t')|$. Thus, a document~$d'$ is considered a candidate source of text reuse for a document~$d$, if at least one of their passage-level hash sets intersects \cite{stein:2019c}: $\exists t \subseteq d$ and $\exists t' \subseteq d' : h(t) \cap h(t')\neq\emptyset$. This means that two documents have to share at least one of the passage-level fingerprint hashes, i.e. have at least one common characteristic, to be deemed a candidate pair.

In our pipeline, the documents are first divided into consecutive passages of $n$~terms, which in turn are each embedded using a bag-of-words representation, i.e., as their term occurrence vector. To find matching passages between two documents, MinHash \cite{broder:97} is chosen as hashing scheme for~$h$. For each passage vector, multiple individual hash functions are applied to each element, and the minimum hash value of each pass is saved, yielding the set minimum hashes for a passage. It can be shown that, for~$m$ individual hashes per passage, two texts with a Jaccard similarity of at least~$m^{-1}$ are guaranteed to produce a hash collision between their hash sets, rendering them a suitable approximation for the lower bound of document similarity \cite{broder:97}. Applying this scheme to all passages in all documents, we detect all document pairs that are locally similar, i.e., which have a word overlap in their bag-of-words passage representations. For example, if documents would be split into~$n=20$ word passages, and~$m=5$ hash fingerprints are calculated per passage, two documents would be identified as pair if they share at least $4$~words across one of the respective passages. A document is never compared to itself.

Hash-based source retrieval allows for a significant reduction of the required computation time, since the hash-based approximation has a linear time complexity with respect to~$|D|$, as opposed to the quadratic complexity of vector comparisons \cite{stein:2019c}. As a result, our source retrieval computation time could be fitted into the allotted budget of two months of computing time on a 130-node Apache Spark cluster, with 12~CPU cores and 196~GB~RAM per node (1560~cores, 250~TB~RAM total). Besides its efficiency, the outlined approach provides us with a second highly beneficial characteristic---the source retrieval step depends on only two parameters: the passage length~$n$, which can be chosen according to computational constraints, and the number of hashes~$m$, which is chosen according to the required minimum similarity separating two passages. By choosing this bound of $m^{-1}$~Jaccard similarity lower than the minimum detection threshold of the subsequent text alignment step, the search space is pruned without a loss in accuracy. Thus, our hash-based similarity search primarily impacts the precision, but not the recall of the source retrieval step with respect to the subsequent text alignment step.

\subsection*{Text Alignment}\label{sub:alignment}

\begin{figure}[ht]
\centering
\includegraphics[width=\textwidth]{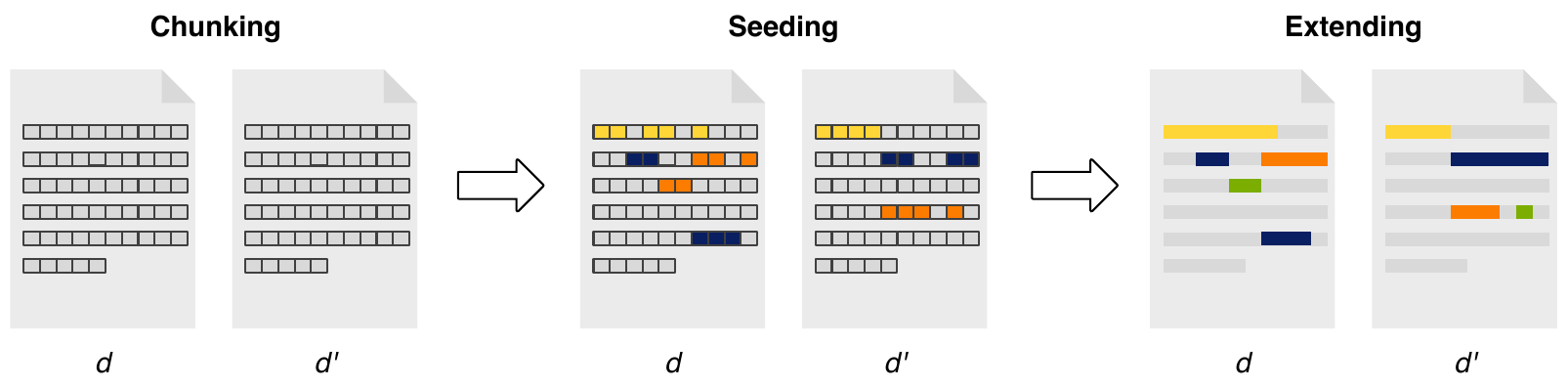}
\caption{Schematic overview of the text alignment process. Each document is divided into chunks, matching chunks are identified between documents, and matches are extended to whole reuse cases.}
\label{fig:alignment}
\end{figure}

Based on the reduced set of comparison candidates~$D_d$ obtained by source retrieval, the text alignment component extracts the exact location of the reused passages of each candidate pair of documents~$d$ and~$d'\ in D_d$. Here we follow the seed-and-extend approach to local sequence alignment \cite{potthast:2012b}. The two-step process is illustrated in Figure~\ref{fig:alignment}. First, both texts are divided into small chunks (\emph{`chunking'}). Then, the matching chunks in the Cartesian product of the two sets of chunks are computed according to a similarity function~$\varphi$. Its purpose is to identify matching textual chunks (`\textit{seeds}`') between two documents~$d$ and~$d'$ that have the same meaning or can be considered instances of the same concept. Finally, sufficiently close matches are combined into larger passages (\emph{`extension'}). Such a pair of passages~$(t \subseteq d, t' \subseteq d')$ in both documents is then output probable reuse case.

With respect to the chunking and seeding steps, two methodological decisions must be made. First, how to divide a given document into small chunks. And second, which similarity function should be used to compare pairs of text chunks to determine whether they have the same meaning. A widely used approach to the former, which has been shown to produce accurate results, is to divide a text into (word)~\emph{n}-grams, i.e., contiguous chunks of words of length~\emph{n} \cite{stamatatos:2011,citron:2015}. These \emph{n}-grams are overlapping and are created by ``sliding'' a window over the text. For example, the sentence ``The quick brown fox jumps over the lazy dog.'' consists of the \emph{4}-grams ``The quick brown fox'', ``quick brown fox jumps'', ``brown fox jumps over'', ``fox jumps over the'', ``jumps over the lazy'', and ``over the lazy dog'' at an overlap of three words. The  \emph{n}-gram chunks of the two documents~$d$ and~$d'$ are then exhaustively compared, i.e., every part of~$d$ is compared to every part of~$d'$. As a similarity function, we again use a hash function, but a string hash function rather than a locale-sensitive function. This choice enables a linear-time comparison of the Cartesian product. Two parameters can be modified in this approach: the \emph{n}-gram size~$n$, and the \emph{n}-gram overlap~$k$.

The matching chunks found through hash collisions indicate ``seeds'' of a potentially longer case of reused text. To determine whether such a case can be found, the \emph{extension} step joins co-aligned matching chunks into longer passages when a sufficient number are found close to each other. In this manner, not only cases of coherent reuse, but also cases where text was copied and then paraphrased can be detected. For example, if consecutive sentences in a source text are copied into a target text, and then another (new) sentence is placed in between them, this would be still considered a case of coherent reuse. Seeding would succeed in identifying the two copied sentences on their own, yet the extension recognizes that both seeds are in close proximity in both documents, and thus outputs a single case of reuse including both. The opposite case is also possible: two chunks from different locations in a source text are placed close to each other in the target texts. Here, again, an extension is required to reconstruct the full scope of reuse. Note that the extension approach depends on a single parameter,~$\Delta$, which is the maximum distance in characters for two seeds in either of a pair of documents below which a single reuse case is assumed \cite{stamatatos:2011}.

We supply a massively parallelizeable implementation of both the source retrieval and text alignment step, which allows for detecting text reuse in the highly scalable manner needed given the amount and length of input documents: 4.2~million publications, totaling 1.1~terabytes of text data. Overall, the text alignment step accounted for an additional 3.5~months of computing time on the aforementioned Spark cluster.

\section*{Data Records}

Two types of data records are included in the \webisReuseCorpus corpus: the reuse case data, which contains all identified cases of text reuse alongside their metadata, and the publication data for each individual document considered when computing the cases, including publication year and field of study annotations. The records can be cross-referenced using a publications' DOI as primary key. The corpus consists of two archive files: \texttt{cases.tar.gz} contains the reuse case data, while \texttt{publications.tar.gz} contains the publication data. Each of these archives bundles multiple partial files in the JSONL format, where every line corresponds to a unique JSON-encoded case or publication. The \webisReuseCorpus corpus is archived at Zenodo (\href{http://doi.org/10.5281/zenodo.5575285}{\texttt{https://doi.org/10.5281/zenodo.5575285}}\cite{zenodo-metadata}). 

Each of the 91,466,374 identified cases of potential text reuse is represented as an individual entry, referencing two different publications. A pair of publications can contain multiple occurrences of text reuse, each of which is treated as a unique case and entry. Each case encompasses two kinds of metadata:
\Ni
\textit{locators}, which identify the matched text by its in-text location, using character offsets to mark start and end, and
\Nii
\textit{context} about the publications involved in the case both by year and by field of study.
In the context of a case, we refer to the first involved publication as~\textit{a}, and to the second as~\textit{b}. This, however, does not indicate a directionality of reuse in terms of publication time. Table~\ref{tab:data-record-cases} gives a detailed overview on all data fields available.

In the publication data, the metadata for all 4,267,166 documents considered for reuse computation is provided. Not every document is involved in at least one reuse case. Publications without detected reuse where included to help contextualize analyses derived from the case data, providing baseline data about distribution of metadata such as fields of study. Table~\ref{tab:data-record-publications} provides detailed information about the fields available for each publication.

\section*{Technical Validation}

This section motivates and details the parameter choices for the source retrieval and text alignment components of the text reuse detection pipeline. For both steps, we strive for maximum accuracy given the constraints for scalability and computational efficiency imposed by the amount of data to be processed. To contextualize the usability of the final dataset, key insights into the distribution of data are given as well.

\subsection*{Source Retrieval}

Objective of the source retrieval step is to prune the search space of document pairs by reducing the number of pairs to be compared in subsequent (computationally expensive) steps. The optimization criterion is recall: Ideally, no document pair containing text reuse should be overlooked, i.e., the false negative rate should be minimized.

To allow for a fine-grained detection of local similarity, the parameters of the source retrieval step are set to~$n=50$ and~$m=10$, i.e., documents are split into passages of 50~words, with 10~MinHash values computed per passage. Consequently, the source retrieval is able to identify pairs of documents that share as little as a 9-word overlap between any two passages. Note that these words do not need to be consecutive since passages are represented in a bag-of-words model. Since the subsequent alignment step operates at the 8-gram level (=~eight consecutive units, see next section), the filtering within the source retrieval step eliminates only pairs for which the subsequent alignment step is guaranteed to not find any matches. Overall, the source retrieval step yields~$3.305\times 10^{12}$ total unique document pairs for further analysis. Given the initial document count of~4,656,302, this represents a pruning of the search space by over~84\% compared to an exhaustive pairwise comparison, rendering the source retrieval step highly effective in improving overall efficiency.

\subsection*{Text Alignment}

The text alignment step identifies the exact location and extent of reuse between two documents. Unlike source retrieval, text alignment is a precision-oriented task: Special emphasis is put on precision over recall for the parameter choice to minimize the number of false positives; a low noise ratio is paramount for meaningful future analyses of the dataset. To ensure the effectiveness of the text alignment, its parameters are chosen by performing a grid-search, using precision, recall, and~$F_{0.5}$ as effectiveness scores. We employ the PAN-13 Text Alignment Corpus for evaluation, which has been previously published as a benchmark dataset for text alignment by the PAN Shared Task on Plagiarism Detection \cite{potthast:2013}. It contains $10,000$~pairs of documents that are subject to different so-called obfuscation strategies that simulate more difficult cases of plagiarism, where the authors tried to hide text-reuse by paraphrasing the copied text to some extent. 

The applied obfuscation strategies include \emph{random obfuscation} (shuffling, adding, deleting, and replacing words or short phrases at random), \emph{cyclic translation obfuscation} (a text is translated into another language, and back to the original language; possibly with more languages in-between), \emph{summary obfuscation} (human-generated summaries of source texts), non-obfuscated plagiarism, and pairs of documents without any text reuse between them. Each of these strategies is equally represented in the corpus with $2,000$~pairs. The grid search identifies an \emph{n}-gram size of~$n = 8$, an \emph{n}-gram overlap of~$k = 7$, and an extension range of~$\Delta = 250$ as optimal. Under these conditions, our implementation of the text alignment step achieves a precision of~0.93 at a recall of~0.46, and thus an~$F_{0.5}$ score of~0.77. This renders our approach highly competitive when compared to other approaches evaluated as part of the PAN Shared Tasks, placing it among the best for precision and~$F_{0.5}$, while being the only approach adhering to the scalability requirements imposed by the scale of our analysis. A full overview of the attained scores for comparison with competing systems is given in Table~\ref{tab:pan13}.

When evaluating separately per obfuscation strategy (Table~\ref{tab:evaluation}), precision is very high throughout, exceeding~0.88 in all cases. This fact places our system within~0.08 for the best-performing, yet computationally more complex approach for each obfuscation strategy. Yet, a drop in recall can be observed for heavily obfuscated text, down to~0.1 for summary obfuscation. This effect is expected and also noticeable among the other approaches presented at~PAN. Moreover, our focus is not on plagiarism but on establishing a general baseline for text reuse in science, where more literal reuse (e.g., citations, idiomatic reuse, template writing, etc.) is expected to be the norm. The heavily obfuscated test sets studied at~PAN were dedicated to study extreme cases of plagiarism, where an author expends much effort to hide the fact via severe forms of paraphrasing. The investment of such an effort, however, goes against the time savings that can be expected from reusing text, so that it can be presumed that the vast amount of paraphrased text reuse will only make smaller changes to a text instead of changing every last \emph{n}-gram. This observation is partially corroborated when reviewing the many cases of academic plagiarism found in dissertation theses throughout the recent years \cite{vroniplag:2021}, where extensive paraphrasing is hardly ever observed. We therefore deemed the investment of an exponentially higher computation cost into retrieving such cases to be uneconomical.

Nevertheless, two measures to tackle this recall issue have been proposed by PAN~participants:
\Ni
customized approaches for each of the different obfuscation strategies, employing heuristics to detect which kind a document pair is exhibiting, and
\Nii
ensemble methods encompassing different seeding and extension strategies, combined into a single result.
However, the first is not applicable to our situation, as there is no ground truth data available to fine-tune such a classification to scientific writing. The second comes at a very high runtime and algorithmic complexity.
This is reflected in Table~\ref{tab:pan13} as well: for each approach, the (asymptotic) algorithmic complexity is noted, for two given sequences of length~$n$ and~$m$. Only six of the approaches presented at~PAN perform in sub-quadratic time, a necessary requirement for large-scale detection. Furthermore, the data specificity for each approach is listed: it denotes how much fine-tuning to the PAN~data has taken place, for example, by crafting specialized corpus-dependent features, using ensemble methods, trained classifiers, or specific approaches for each of the different obfuscation strategies, all of which reduce the generalizability and transferability of the approaches to other data. 

Out of the four other approaches to combine sub-quadratic runtime complexity with low data specificity, ours performs best with regard to recall and the $F_{0.5}$~score, and is second in precision by a very close margin. Against this background, our achieved detection performance is comparably outstanding, given the specialized requirements in terms of scalability as well as the focus on a high-precision identification.

\subsection*{Dataset Properties}

\begin{figure}[ht]
\centering
\includegraphics[width=\textwidth]{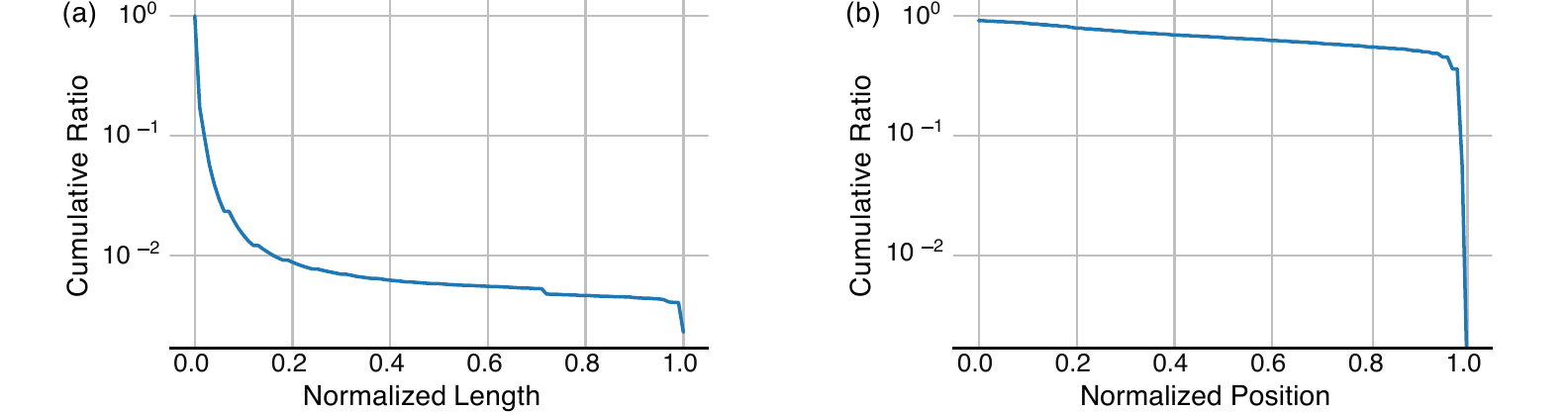}
\caption{\textbf{(a)} cumulative ratio of reuse cases by normalized case length. \textbf{(b)} cumulative ratio of reuse cases by normalized position.}
\label{fig:distributions}
\end{figure}

Table~\ref{tab:disciplines} lists the distribution of identified cases and original publications across scientific disciplines. The relative share of disciplines is approximately the same for both, with only natural sciences having a decreased share in cases compared to publications. Figure~\ref{fig:distributions} shows cumulative ratios of cases per normalized length, i.e., matched case length divided by total publication length, and normalized position, i.e., start offset divided by total publication length. Less than one percent of cases encompasses more than~20\% of the original publication. 90\%~of cases cover at most~1\% of the original publication. Some duplicate publications were contained in the dataset under different~DOIs, as made evident by the small spike at normalized length larger than~0.99. Most of the reuse cases occur in the last~5\% of a publication. This is likely due to author contribution statements, copyright notices, or acknowledgements contributing a majority of boilerplate text reuse. The distribution over the rest of the relative positions is approximately uniform. Overall, these three key properties highlight the validity of the data, and, combined with the overall amount of cases, this allows to build focused subsets of substantial size for downstream tasks and analysis.

\section*{Usage Notes}

\subsection*{Applications and Examples}

Based on the various reuse cases contained in \webisReuseCorpus, a variety of research questions can be answered. These include the study of discipline-specific writing practices, comparative studies, and a variety of machine learning tasks. The corpus includes a wide range of reuse cases, and the operationalization of text reuse is deliberately chosen to encompass many phenomena of reuse. \webisReuseCorpus contains mainly ``innocent'' instances of reuse, such as short phrases reused by the same authors, standardized technical language, wording prescribed by publishers (e.g., licenses), or established wording of recurring blocks of text (e.g., author contributions). In particular, the corpus is not a collection of plagiarism cases, and we refrain from judging the legitimacy of the cases. 

To illustrate the variety of reuse cases, Table~\ref{tab:example-cases} shows three examples, each representing a different type of legitimate reuse. The first is a quotation, where two authors each quote verbatim from a tertiary source. The second is an example of a paraphrase and comes from two different articles by the same author on the same topic. The third is an example of reuse of boilerplate text, as both are contribution statements whose structure is highly formalized, the difference being the different contributions of the various authors. Because of the many types of reuse, answering a particular research question may require filtering techniques to create a subset suitable for the application at hand. The metadata included allows for fine-grained filtering, e.g., by location in the text, extent, time, and specific scientific fields. 

\subsection*{Data Accessibility}

The dataset is distributed as JSONL~files of reuse case metadata to enable efficient streaming and filtering even with low computational resources. Since some analyses require the text of the reuse cases \webisReuseCorpus, researchers can hydrate the corpus using the text preprocessing component of our processing pipeline, which is included with the dataset as a standalone Python script. This enables the conversion of GROBID extraction results, e.g. from the CORE repository, into a compatible text format so that the locators contained in the dataset can be used to recover the text portions of individual reuse cases. Since the publications the corpus is based upon are open access, the original PDF~files can be easily retrieved by their~DOI, further lowering the barrier of access. We also provide the code used to compute the corpus statistics in this article to give others interested in working with the data an example of use.

\subsection*{Ethical Considerations}

Our dataset includes contemporary scientific texts (``papers'') with the goal of examining the occurrence, nature, and types of text reuse that result from scientific writing practices. Given general ethical considerations for datasets \cite{peng:2021}, three are particularly relevant to the proposed collection:
\Ni
privacy of the individuals included in the data, 
\Nii
effects of biases on downstream use, and
\Niii
dataset usage for dubious purposes.
We therefore took into account a consensus on best-practices for ethical dataset creation \cite{mieskes:2017,leidner:2017,gebru:2021}.

Ad~(1).
While the corpus does not contain author names or identifiers, all included papers are freely accessible, so it is possible to determine the identity of individual authors by referring to the original CORE/OAG~data or simply by accessing a~DOI. However, we do not consider this to be problematic, since all those who participate professionally in scientific discourse agree, by publishing their contributions, that they will go down in the scientific annals under their name and that they can be examined by anyone. This is especially true for articles under an open access license, where consent to create derivative works, public archiving, and exploitation is implied.

Ad~(2). 
Two types of bias can arise from our process of dataset curation. First, by using only open access publications, the types, disciplines, and characteristics of the papers included may not be representative of science as a whole. Since it is not yet possible to analyze all scientific papers ever published, we try to minimize this risk by using as large a sample as possible. Second, the operationalization and implementation of text reuse may vary for more specific research questions on this topic. Our goal was therefore to employ a very general and inclusive detection approach. For downstream tasks, some of the cases included are not interesting, and the data can filtered to obtain a more targeted collection. Moreover, by reproducibly documenting its creation process, we aim to maximize its extensibility in the future.

Ad~(3).
We estimate the potential for misuse of the dataset to be low. One contentious issue is the use of the dataset to broadly target academics on their writing practices. However, this has not been done with previously published text reuse datasets. The novelty of this dataset compared to previous datasets is the scope, from which no new doubtful use cases emerge. Furthermore, the dataset explicitly refrains from classifying the legitimacy of reuse cases. Both the operationalization of the term and its intended use are intended to examine all types and techniques of reuse of text in science, not plagiarism in particular.

\section*{Code Availability}

The complete source code used for candidate retrieval and text alignment is openly accessible and permanently available on GitHub (\url{https://github.com/webis-de/arxiv21-stereo-scientific-text-reuse}). The data processing pipeline is written in Python~3.7, utilizing the \texttt{pyspark} framework. The compute cluster on which we carried out the data processing and our experiments run Spark Version~2.4.8. The text alignment component is written in Go~1.16 and can be used as a standalone application. Detailed documentation about each pipeline component, recommendations for compute resources, and suggestions for parameter choices are distributed alongside the code to facilitate code reuse.


\section*{Acknowledgments}

This work was supported by grant no. 01PW18015B of the German Federal Ministry of Education and Research.

\section*{Author Contributions Statement}

L.G., M.P., W.K. and B.St.\ conceived the experiment. L.G. conducted the experiment. W.K. and B.Si.\ assembled the metadata and carried out the plain text extraction. All authors wrote and reviewed the manuscript.

\section*{Competing Interests}

The authors declare no competing interests.

\newpage
\section*{Figures \& Tables}

\begin{table}[ht]
\centering
\small
\caption{Data record overview for case data. \textit{Key} denotes the top-level JSON key by which fields are identified, \textit{Description} provides an explanation of the contained data, and \textit{Type} denotes the data type; ``optional'' indicates fields that can be empty. The text fields are included only in the full version of the corpus.}
\label{tab:data-record-cases}
\renewcommand{\tabcolsep}{15pt}
\hspace*{-2em}
\begin{tabular}{@{}c@{}c@{\quad}lll@{}}
\toprule
& & \textbf{\textsf{Key}} & \textbf{\textsf{Description}} & \textbf{\textsf{Type}} \\
\midrule
& & \texttt{id} & Unique identifier for this case, in UUID format. & String \\
\midrule
\multirow{8}{*}{\vspace*{-5.2pt}$\left.\rotatebox[origin=c]{90}{\hbox to 24ex{\textit{\textsf{\small\hfill Publication A\hfill}}}}\right\{$}
& \multirow{3}{*}{\rotatebox[origin=c]{90}{\textit{\textsf{\small Locator}}}}
& \texttt{begin\_a} & Start location of matched text, measured as character offset.& Integer\\
& & \texttt{end\_a} & End location of matched text, measured as character offset.& Integer\\
& & \texttt{doc\_length\_a} & Total length of publication A in characters. & Integer\\
\cmidrule{2-5}
& \multirow{5}{*}{\rotatebox[origin=c]{90}{\textit{\textsf{\small Context}}}}
& \texttt{doi\_a} & DOI identifier for publication A.& String\\
& & \texttt{year\_a} & Publication year for publication A.& Integer (optional)\\
& & \texttt{field\_a} & Field(s) of study for publication A.& String Array (optional)\\
& & \texttt{area\_a} & Scientific area(s) for publication A. & String Array (optional)\\
& & \texttt{discipline\_a} & Scientific discipline(s) for publication A.& String Array (optional)\\[1.0ex]
\midrule
\multirow{8}{*}{\vspace*{-5.2pt}$\left.\rotatebox[origin=c]{90}{\hbox to 24ex{\textit{\textsf{\small\hfill Publication B\hfill}}}}\right\{$}
& \multirow{3}{*}{\rotatebox[origin=c]{90}{\textit{\textsf{\small Locator}}}}
& \texttt{begin\_b} & Start location of matched text, measured as character offset.& Integer\\
& & \texttt{end\_b} & End location of matched text, measured as character offset.& Integer\\
& & \texttt{doc\_length\_b} & Total length of publication B in characters. & Integer\\
\cmidrule{2-5}
& \multirow{5}{*}{\rotatebox[origin=c]{90}{\textit{\textsf{\small Context}}}}
& \texttt{doi\_b} & DOI identifier for publication B.& String\\
& & \texttt{year\_b} & Publication year for publication B.& Integer (optional)\\
& & \texttt{field\_b} & Field(s) of study for publication B.& String Array (optional)\\
& & \texttt{area\_b} & Scientific area(s) for publication B. & String Array (optional)\\
& & \texttt{discipline\_b} & Scientific discipline(s) for publication B.& String Array (optional)\\[1.0ex]
\bottomrule
\end{tabular}
\end{table}

\begin{table}[ht]
\centering
\small
\caption{Data record overview for publication data. \textit{Key} denotes the top-level JSON key fields are identified by, \textit{Description} provides an explanation of the contained data, and \textit{Type} denotes the data type; ``optional'' indicates fields that can be empty.}
\label{tab:data-record-publications}
\renewcommand{\tabcolsep}{10pt}
\begin{tabular}{@{}c@{\quad}lll@{}}
\toprule
& \textbf{\textsf{Key}} & \textbf{\textsf{Description}} & \textbf{\textsf{Type}} \\
\midrule
\parbox[t]{0.7em}{\multirow{6}{*}{\rotatebox[origin=c]{90}{\textit{\textsf{\small Metadata}}}}}
& \texttt{doi}         & DOI identifier of the publication.             & String\\
& \texttt{doc\_length} & Total length of the publication in characters. & Integer\\
& \texttt{year}        & Publication year of the publication.           & String Array (optional)\\
& \texttt{field}       & Field of study of the publication.             & String Array (optional)\\
& \texttt{area}        & Scientific area of the publication.            & String Array (optional)\\
& \texttt{discipline}  & Scientific discipline of the publication.      & String Array (optional)\\
\bottomrule
\end{tabular}
\end{table}

\begin{table}[ht]
\setlength{\tabcolsep}{4pt}
\caption{Precision, Recall, and $F_{0.5}$ score of competing alignment approaches by teams participating at the PAN Shared Tasks, taken from Potthast et al. (2013)\cite{potthast:2013} and Potthast et al. (2014)\cite{potthast:2014} and sorted descending by precision. The listed approaches are described in detail in the PAN Workshop proceedings \cite{pan:2012,pan:2013,pan:2014}. Complexity denotes the runtime complexity of the approach; Data Specificity denotes the degree of PAN-specific optimizations reducing transferability to other data domains, i.e. ensemble methods, trained approaches, feature extraction, or special handling of corpus characteristics.}
\label{tab:pan13}
\centering
\begin{tabular}{@{}l*{3}{c}cc@{}}
\multicolumn{6}{c}{\textsf{\textbf{PAN13 Evaluation Corpus}}} \\[1.1ex]
\toprule
\textsf{\textbf{Team}}  & \textsf{\textbf{Precision$\downarrow$}} & \textsf{\textbf{Recall }}       & \textsf{\textbf{F\textsubscript{0.5}}}         & \textsf{\textbf{Complexity}}    & \textsf{\textbf{Data Specificity}}  \\
\midrule
        Glinos (2014)   &       0.96            &    0.79       &  0.92             & $O(nm)$       & Medium            \\
       Jayapal (2012)   &       0.95            &    0.22       &  0.57             & $O(n+m)$      & Low               \\
       Nourian (2013)   &       0.95            &    0.43       &  0.76             & ---           & ---               \\
         Gross (2014)   &       0.93            &    0.77       &  0.89             & $O(nm)$       & Low               \\
         Alvi (2014)    &       0.93            &    0.55       &  0.82             & $O(n+m)$      & Medium            \\
        \textbf{Ours}   &\textbf{0.93}          &\textbf{0.46}  &  \textbf{0.77}    & $O(n+m)$      & Low               \\
      Baseline (2014)   &       0.93            &    0.34       &  0.69             & $O(n+m)$      & Low               \\
    Palkovskii (2014)   &       0.92            &    0.83       &  0.90             & $O(nm)$       & High              \\
      Torrejon (2014)   &       0.90            &    0.77       &  0.87             & $O(nm)$       & High              \\
        Gillam (2014)   &       0.89            &    0.17       &  0.48             & $O(nm)$       & High              \\ %
    Oberreuter (2014)   &       0.89            &    0.86       &  0.88             & ---           & ---               \\
    Oberreuter (2012)   &       0.89            &    0.77       &  0.86             & ---           & ---               \\ %
        Gillam (2012)   &       0.89            &    0.27       &  0.61             & $O(nm)$       & High              \\ %
      Torrejon (2013)   &       0.89            &    0.76       &  0.86             & $O(nm)$       & High              \\ %
        Gillam (2013)   &       0.88            &    0.26       &  0.60             & $O(nm)$       & High              \\ %
       Jayapal (2013)   &       0.88            &    0.38       &  0.70             & $O(n+m)$      & Low               \\
 Sanchez-Perez (2014)   &       0.88            &    0.88       &  0.88             & $O(nm)$       & High              \\
      Kueppers (2012)   &       0.87            &    0.51       &  0.76             & $O(nm)$       & Low               \\
      Shrestha (2013)   &       0.87            &    0.74       &  0.84             & $O(nm)$       & Medium            \\
        Saremi (2013)   &       0.87            &    0.77       &  0.85             & ---           & ---               \\
      Shrestha (2014)   &       0.86            &    0.84       &  0.86             & $O(nm)$       & Medium            \\
          Kong (2012)   &       0.85            &    0.82       &  0.84             & $O(nm)$       & Low               \\
      Suchomel (2012)   &       0.84            &    0.65       &  0.79             & $O(n+m)$      & Medium            \\
          Kong (2014)   &       0.84            &    0.81       &  0.83             & $O(nm)$       & Low               \\
          Kong (2013)   &       0.83            &    0.81       &  0.83             & $O(nm)$       & Low               \\
      Torrejon (2012)   &       0.83            &    0.75       &  0.81             & $O(nm)$       & High              \\ %
    Palkovskii (2013)   &       0.82            &    0.54       &  0.74             & $O(nm)$       & High              \\
    Palkovskii (2012)   &       0.82            &    0.76       &  0.81             & $O(nm)$       & High              \\
         Abnar (2014)   &       0.77            &    0.61       &  0.73             & $O(nm)$       & Medium            \\
      Suchomel (2013)   &       0.73            &    0.77       &  0.74             & $O(n+m)$      & Medium            \\
  Sanchez-Vega (2012)   &       0.40            &    0.56       &  0.42             & $O(nm)$       & Medium            \\
\bottomrule
\end{tabular}
\end{table}

\begin{table}[ht]
\centering
\caption{Precision, Recall, and $F_{0.5}$ of the text alignment component per obfuscation strategy and on the complete evaluation corpus.}
\label{tab:evaluation}
\begin{tabular}{lccc}
\multicolumn{4}{c}{\textsf{\textbf{PAN13 Evaluation Corpus}}} \\[1.1ex]
\toprule
                        &  \textsf{\textbf{Precision}} &  \textsf{\textbf{Recall}} &\textsf{\textbf{F\textsubscript{0.5}}} \\
\midrule
No Obfuscation          &       0.88 &    0.90 &  0.88 \\
No Plagiarism           &       1.00 &    1.00 &  1.00 \\
Random Obfuscation      &       0.90 &    0.11 &  0.37 \\
Summary Obfuscation     &       0.99 &    0.10 &  0.36 \\
Translation Obfuscation &       0.88 &    0.16 &  0.46 \\ \midrule
Entire Corpus           &       0.93 &    0.46 &  0.77 \\
\bottomrule
\end{tabular}
\end{table}

\begin{table}[ht]
\caption{Number and ratio of cases and publications per scientific discipline. Percentages can exceed 100 in sum due to multiple membership.}
\label{tab:disciplines}
\centering
\begin{tabular}{lrrrr}
\toprule
  \textsf{\textbf{Discipline}}  & \multicolumn{2}{c}{\textsf{\textbf{Cases}}} & \multicolumn{2}{c}{\textsf{\textbf{Publications}}} \\
\midrule
  Natural Sciences              & 21,504,070 &          ({\color{gray} 24\%}) & 1,606,599 &                  ({\color{gray} 38\%}) \\
  Engineering Sciences          & 14,753,613 &          ({\color{gray} 16\%}) &   911,226 &                  ({\color{gray} 21\%}) \\
  Life Sciences                 & 43,358,077 &          ({\color{gray} 47\%}) & 1,646,843 &                  ({\color{gray} 39\%}) \\
  Humanities \& Social Sciences & 15,265,777 &          ({\color{gray} 17\%}) &   748,298 &                  ({\color{gray} 18\%}) \\
  Total                         & 91,466,374 &                                & 4,267,166 &                                        \\
\bottomrule
\end{tabular}
\end{table}

\begin{table}[ht]
\caption{Matched text with before and after context for selected reuse examples. Portions with highest similarity highlighted. DOIs of original publications given.}
\label{tab:example-cases}
\centering
\begin{tabular}{@{}p{.49\textwidth}p{.49\textwidth}@{}}
\toprule
\multicolumn{1}{c}{\textsf{\textbf{Publication A}}} & \multicolumn{1}{c}{\textsf{\textbf{Publication B}}} \\ \midrule

[\dots] children than men in lower classes nor experienced the same downward mobility. Clark writes, \ctext[RGB]{255,214,75}{``Thus we may speculate that England's advantage lay in the rapid cultural, and potentially also genetic, diffusion of the values of the economically successful throughout society in the years 1200-1800'' (p. 271).} It does not serve the book well to dwell on the rebuttal of its evolutionary [\dots] \newline {\scriptsize\color{gray}{10.1111/j.1475-4991.2009.00354.x}}
& [\dots] Impact on the standard of living, but eventually they led to the end of the long Malthusian era. \ctext[RGB]{255,214,75}{``Thus we may speculate that England's advantage lay in the rapid cultural, and potentially also genetic, diffusion of the values of the economically successful throughout society in the years 1200-1800'' (p. 271).} Finally, Clark considers the great divergence among today's economies. Because [\dots] \newline {\scriptsize\color{gray}{10.1111/j.1468-0289.2008.00432\_10.x}}
\\
\midrule

[\dots] static approximation of the ColourSinglet Model, we have seen that the production amplitude \ctext[RGB]{255,214,75}{receives contributions from two different cuts. The first one in its static limit gives the colour-singlet mechanism. The second one has not been considered so far. We treat it in a gauge-invariant manner by introducing necessary new 4-point vertices, suggestive of the colour-octet mechanism.} This new contribution can be as large as the colour-singlet mechanism at high \newline {\scriptsize\color{gray}{10.1063/1.2122163}}
& [\dots] and we show that the lowest-order mechanism for heavy-quarkonium production \ctext[RGB]{255,214,75}{receives in general contributions from two different cuts. The first one corresponds to the usual colour-singlet mechanism. The second one has not been considered so far. We treat it in a gauge-invariant manner, and introduce new 4-point vertices, suggestive of the colour-octet mechanism}. These new objects enable us to go beyond the static approximation. We show that the contribution of [\dots] \newline {\scriptsize\color{gray}{10.1016/j.physletb.2005.11.073}}
\\
\midrule

[\dots] \ctext[RGB]{255,214,75}{The authors declare that they have no competing interests.} MS and MMD drafted and wrote the manuscript. MS, MMD and KCRP participated in the care of the patient and interpretation of the investigations. \ctext[RGB]{255,214,75}{All authors read and approved the final manuscript.} [\dots] \newline {\scriptsize\color{gray}{10.1186/1757-1626-2-147}}
& [\dots] \ctext[RGB]{255,214,75}{The authors declare that they have no competing interests.} Authors' contributions: AJB is the chief investigator of the CASP trial, he is responsible for the conduct of the study and the clinical supervision and training of the therapists and he wrote the first draft of this paper. CS drafted an earlier version of the grant proposal and collaborated with AJB, MT, RR \& PH to produce the successful proposal. CS also provided clinical supervision and therapist training. LS edited and revised the proposal and is responsible for the day to day running of the trial. \ctext[RGB]{255,214,75}{All authors read and approved the final manuscript.} [\dots] \newline {\scriptsize\color{gray}{10.1186/1471-244X-13-199}} \\
\bottomrule
\end{tabular}
\end{table}

\end{document}